\documentclass[a4paper]{jpconf}

\usepackage[latin1]{inputenc}
\usepackage{epsfig}
\usepackage[english,french]{babel}
\usepackage{color}
\usepackage{graphicx}
\usepackage{dcolumn}
\usepackage{moreverb}
\usepackage[all]{xy}
\usepackage{amsmath,amssymb,amsfonts}

\usepackage{amscd}

\begin{document}
\title{Higher order superintegrability, Painlev\'e transcendents and
representations of polynomial algebras}

\author{Ian Marquette}

\address{School of Mathematics and Physics, The University of Queensland, 4072 QLD Australia}

\ead{i.marquette@uq.edu.au}

\begin{abstract}
In recent years, progress toward the classification of superintegrable systems with higher order integrals of motion has been made. In particular, a complete classification of all exotic potentials with a third or a fourth order integrals, and allowing separation of variables in Cartesian coordinates. All doubly exotic potentials with a fifth order integral have also been completely classified. It has been demonstrated how the Chazy class of third order differential equations plays an important role in solving determining equations. Moreover, taking advantage of various operator algebras defined as Abelian, Heisenberg, Conformal and Ladder case of operator algebras, we re-derived these models. These new techniques also provided further examples of superintegrable Hamiltonian with integrals of arbitrary order. It has been conjectured that all quantum superintegrable potentials that do not satisfy any linear equation satisfy nonlinear equations having the Painlev\'e property. In addition, it has been discovered that their integrals naturally generate finitely generated polynomial algebras and the representations can be exploited to calculate the energy spectrum. For certain very interesting cases associated with exceptional orthogonal polynomials, these algebraic structures do not allow to calculate the full spectrum and degeneracies. It has been demonstrated that alternative sets of integrals which can be build and used to provide a complete solution. This this allow to make another conjecture i.e. that higher order superintegrable systems can be solved algebraically, they require alternative set of integrals than the one provided by a direct approach.
\end{abstract}

\section{Introduction}

The connection between the n-dimensional Kepler-Coulomb and harmonic oscillator and Lie algebras $su(n)$ and $so(n+1)$ have been central to the development of algebraic techniques in physics and study of exactly solvable models \cite{Fo,Ba,Hil,Sud1,Ban1,Lou1,Ras1,Bak1,Mos1,Lou2,Bud1,Bar1,Fra1,Hwa1}. They also motivated application of finite dimensional uunitary representations and Casimir operators and development of algebraic methods to study Hamiltonians. These two models are exactly solvable but as well superintegrable. A systematic search for two-dimensional superintegrable Hamiltonians with two second order integrals was started in the 60ties by Winternitz \cite{Win65}.  Many of their properties have been discovered for these models such as multiseparability, exact solvability and degenerate spectrum. Over the years, a complete classification of quadratically superintegrable systems on two-dimensional conformally flat space was performed \cite{Mil13}. Progress has been reported for the three dimensional case \cite{Mil17}.  

A program to study superintegrable systems with a second order integrals and another higher order integrals of Nth order is more recent and has been started fifteen years ago. A first results consisted in obtaining the determining equations (N=3) from a third order integral \cite{Gra02} and classifying all Hamiltonians allowing separation of variable in Cartesian coordinates with a third order integrals \cite{Gra04}. A connection with Painlev\'e transcendents was also established for the first time and demonstrated the significance of studying higher order superintegrability.

Much recently, it has been recognized how the search can be narrowed to exotic potentials i.e. the potential do not satisfy any linear differential equation. All exotic potentials for integrals of fourth order (N=4) \cite{Saj17} and doubly exotic systems with integral of fifth order (N=5) \cite{Ism18} were obtained. These potentials can be rewritten in term of the first, second, third, fourth and fifth Painlev\'e transcendents. Superintegrable systems allowing separation of variables in parabolic coordinates for N=3 \cite{Pop12}, as well as systems allowing separation of variables in polar coordinates for N=3 \cite{Tre10} and N=4 \cite{Adr17,Adr18a}. Models were obtained in terms of the sixth Painlev\'e transcendents. Some work have been done in regard of the case of Nth order integrals related to superintegrable systems with separation of variables in Cartesian coordinates \cite{Sno15} and \cite{Pos15}. In particular various alternative form for the integrals of motion have been obtained. Similar work for Nth order integrals have also been done in regard of polar coordinates \cite{Adr18b}. 

Recently alternative approach based on one-dimensional operator algebras have been used to obtained new superintegrable systems allowing separation of variable in Cartesian coordinates with integrals of arbitrary order \cite{Saj18}. Systems involving Painlev\'e transcendents with integrals of arbitrary order have been generated.

These results had lead to formulate a conjecture that for all maximally superintegrable systems with exotic potentials in quantum mechanics the corresponding nonlinear differential equations possess the Painlev\'e property \cite{Mar18}.

Another program has been started in regard of superintegrable systems which is to construct their symmetry algebra and calculate their energy spectrum algebraically via representations. It has been discovered more than 25 years ago \cite{Gra1,Das1} that the integrals of motion generate naturally quadratic algebra and in particular the case of the Racah algebra \cite{Vin3}. Since they have been applied widely to obtain energy spectrum of superintegrable systems. It has been observed that higher rank quadratic algebra can be exploited \cite{Faz15,Bie1,Li18,Tan1} as well. Example of polynomial algebras find applications \cite{Bon1,Kal12,Que15,Isa1} and in particular the cubic algebras \cite{mar09a,mar09b} in regard of fourth Painlev\'e transcendent models. A puzzling phenomena has been noticed \cite{mar09b} consisting in incomplete algebraic description of the spectrum and the degeneracies for certain values of the parameters of the fourth Painlev\'e transcendent.

The story of that problem took an interesting turn as it has been shown that it was connected \cite{mar13a,mar13b,mar14,mar16} with  exceptional orthogonal polynomials \cite{cq09,gomez10a,gomez09,gomez10b,oda11,oda13,ull14,ull13}.  New integrals can be created via Krein-Adler and Darboux-Crum for k-step extention models and allowed as well a complete algebraic description of the levels and degeneracies of the fourth Painlev\'e transcendent Hamiltonian. 

The purpose of this paper is two fold. First, this paper intend to review recent results on the classification of superintegrable systems \cite{Saj17,Saj18,Mar18}, Painlev\'e property of quantum superintegrable and connection with Chazy class of equations. Secondly, we will describe algebraic solution for the fourth Painleve transcendent via results on Darboux-Crum and Krein-Adler chains. The paper is organized as follow in section 2, we recall results on the quadratic and more generally polynomial algebras. In section 3, we review in particular classification using a direct approach for N=3, N=4 and cubic algebra in one example. In Section 4, we review one dimensional operator algebra method introduced recently. In Section 5, we present discussion of how ladder operator can be factorized in term of supercharge which allow to obtain the wavefunction for case in term of fourth and fifth Painleve and apply to an example together with cubic algebra of section 4 to point out the incomplete solution. In section 6, we discuss how ladder operator are in fact not unique and one can find for a given Hamiltonian alternative ladder which allow
to obtain alternative set of integrals.

\section{Polynomial algebras and superintegrable systems}

An Hamiltonian system on a n-dimensional space ( generally on Riemannian manifold ) with Hamiltonian H 
\begin{equation*}
H=\frac{1}{2}g^{ik}p_{i}p_{k}+V(\vec{x})
\end{equation*}
is integrable if it allows n integrals of motion that are well defined, in involution $\{H,X_{a}\}_{p}=0$, $\{X_{a},X_{b}\}_{p}=0$, a,b=1,...,n-1 and functionally independent. A system is superintegrable if it admits $n+k$ (with $k = 1,...,n-1$) functionally independent constants of the motion (well defined). Maximally superintegrable if $k=n-1$. In quantum mechanics $\{H,X_{a},Y_{b}\}$ are well defined quantum mechanical operators and form an algebraically independent set.

\begin{equation*}
H=\frac{1}{2}\vec{p}^{2}+V(x,y) 
\end{equation*}
was considered and two integrals admit the general form
\[A=\sum_{i,k=1}^{2}\{f^{ik}(x,y),p_{i}p_{k}\}+\sum_{i=1}^{2}g^{i}(x,y)p_{i}+\phi(x,y),j=1,2. \]
\[B=\sum_{i,k=1}^{2}\{v^{ik}(x,y),p_{i}p_{k}\}+\sum_{i=1}^{2}w^{i}(x,y)p_{i}+\psi(x,y),j=1,2. \]

The quadratic algebra take the form, it includes the Racah algebra as a special case

\begin{subequations}\label{quadalg}
\begin{align}
[A,B]&=C \label{quadalga}\\
[A,C]&=\alpha A^{2}+\gamma \{A,B\} +\delta A +\epsilon B + \zeta \label{quadalgb} \\
[B,C]&=a A^{2} -\gamma B^{2} -\alpha \{A,B\} + d A - \delta B + z. \label{quadalgc} \\
\end{align}
\end{subequations}

It admit one Casimir operator which is cubic in the generator

\begin{align}\label{quadcas}
K&=C^{2}-\alpha\{A^{2},B\}-\gamma\{A,B^{2}\}+(\alpha \gamma -
\delta)\{A,B\}+(\gamma^{2}-\epsilon)B^{2} \\ \nonumber
+&(\gamma \delta - 2\zeta)B +\frac{2 a}{3}A^{3}+(d+\frac{a \gamma}{3}\alpha^{2})A^{2}+(\frac{a \epsilon}{3}+\alpha \delta +2z)A. \\ \nonumber
\end{align}

It has been shown it can be exploited to obtain finite dimensional unitary representations of the quadratic algebras. Recent results have shown the importance of that algebra in the classification of two dimensional quadratically superintegrable system on conformally flat spaces. The complex spaces admitting at least three 2nd order symmetries, flat space, complex 2-sphere, the four Darboux spaces, eleven 4 parameter Koenigs spaces. There are 59 2nd order superintegrable systems in 2D, under the Stackel transform, the systems divide into 12 equivalence classes.  6 with nondegenerate 3-parameter potentials (S9,E1,E2,E30,E8,E10), 6 with degenerate 1-parameter potentials (S3,E3,E4,E5,E6,E14), all these systems are related via process of contraction of quadratic algebra related to e(2,C) and o(3,C) along the idea of Wigner-Inonu. These contraction allow to related to the full askey scheme of orthogonal polynomials. More generally Bocher contraction have been identified \cite{Kal13,Hei15,Esc17}.

Algebraic derivation of spectrum using finite dimensional unitary representation has been extended to classes of superintegrable systems that decomposes into direct sums of higher rank Lie algebras and quadratic algebras. Their structure constants also depend on central element or even on elements in the centre of the universal enveloping algebra of some Lie algebra. Recently, an embedded type of structure has been discovered for a n-dimensional superintegrable system and even an extension of that higher rank quadratic algebra with structure constant that depend on Casimir operator of a Lie algebra \cite{Zhe18}. These quadratic algebras of superintegrable systems, are still poorly understood, as their classification remain an open problem as well as systematic study of their representations and properties of their universal enveloping algebra and its centre remains to be obtained.In recent years, the construction of Casimir operator as well as realization as deformed oscillator algebra have been pursue for polynomial algebra generated from underlying second and Nth order integrals of motion \cite{Isa1}

\begin{subequations}\label{polalg}
\begin{align}
[A,B]&=C \label{polalga}\\
[A,C]&=\sum_{i=1}^{\left \lfloor \frac{N}{2}+1 \right \rfloor }\alpha_{i}A^{i}+\delta B +\epsilon + \beta \{A,B\} \label{polalgb} \\
[B,C]&=\sum_{i=1}^{N}\lambda_{i}A^{i}+\rho B^{2} +\eta B + \sum_{i}^{\left \lfloor \frac{N}{2}\right \rfloor}\omega_{i}\{A^{i},B\}+\zeta \label{quadalgc} \\
\end{align}
\end{subequations}

Further constraints are imposed from the Jacobi identity on the structure constants. Various commutator and anti commutator identities were established and as well as recurrence relations. Another algebraic structure which play a role in superintegrability is the deformed oscillator algebras \cite{Das1} $\{b^{\dagger},b,N\}$ 

\begin{subequations}\label{oscalg}
\begin{align}
[N,b^{\dagger}]&=b^{\dagger}\label{oscalga}\\
[N,b]&=-b  \label{oscalgb} \\
b^{\dagger}b&=\Phi(N) \label{oscalgc} \\
bb^{\dagger}&=\Phi(N+1) \label{oscalgd} \\
\end{align}
\end{subequations} 

It has been demonstrated that realization as deformed oscillator algebra exist for polynomial algebra of arbitrary order generated by integrals
of order 2 and N of the form $A=A(N)$ and $B=b(N)+b^{\dagger}\rho(N)+\rho(N)b$. Explicit expression have been obtained for the quadratic \cite{Das1},  cubic \cite{mar09a,mar09b} and quartic case \cite{mar12}. Let us consider the cubic case, which take the form

\begin{subequations}\label{cubalg}
\begin{align}
[A,B]&=C \label{cubalga}\\
[A,C]&=\alpha A^{2} + \beta \{A,B\} + \gamma A + \delta B + \epsilon \label{cubalgb} \\
[B,C]&=\mu A^{3} + \nu A^{2} - \beta B^{2} - \alpha \{A,B\} + \xi A -
\gamma B + \zeta \label{cubalgc} \\
\end{align}
\end{subequations}

The Casimir operator

\begin{align}\label{cascub}
K =& C^{2} - \alpha \{A^{2},B\} - \beta \{A,B^{2}\} + (\alpha
\beta - \gamma)\{A,B\} + (\beta^{2} - \delta)B^{2}\\
+&(\beta \gamma - 2\epsilon)B+\frac{\mu}{2}A^{4} +
\frac{2}{3}(\nu+\mu \beta)A^{3}+(-\frac{1}{6}\mu \beta^{2} +
\frac{\beta \nu}{3} + \frac{\delta \mu}{2} + \alpha^{2} + \xi)A^{2}\\
+&(-\frac{1}{6}\mu \beta \delta + \frac{\delta \nu}{3} + \alpha
\gamma + 2\zeta)A
\end{align}

Two types of realization have been constructed (i.e. for $\beta=0$ and $\beta \neq 0$ ) and explicit formula for $A(N)$, $b(N)$, $\rho(N)$
and $\Phi(N)$ have been provided. The structure function $\Phi(N)$ take the form of polynomials of order 4 and 10 with coefficients involving the structure constants of the cubic algebra. These structure constants in general may depend on parameters of the quantum model such a coupling constant, but can involve the Hamiltonian itself. The algebraic method requires to write the Casimir operator in term of the central elements only which can by finding factorized form to the integrals of motion or using advantage of presence inner structures to the integrals in term of supercharges, ladder or shift operators for which the action on wave function take simpler form. In the case of polynomial algebras, contrary to the quadratic case, it has been observed that there are cases for which these algebraic approaches provide the spectrum, but an incomplete description of the deneneracies or even missing levels. These cases are exactly the one related to fourth Painlev\'e transcendent and specific values of the parameter for which the solution in term of generalized Hermite and Okamoto in fact take the form of exceptional orthogonal polynomials of Hermite of type III.

\section{N=3 and N=4: Exotic potential and Painlev\'e transcendents}

In this section, we will review superintegrable systems on two dimensional Euclidean space separable in Cartesian coordinates with
a second and additional integrals which is a third or a fourth order integral of motion. We will discuss how they are connected with Painlev\'e transcendents \cite{In56,Pa02,Fuchs84,Gam10} and more generally the Chazy class of nonlinear differential equations and their reduction to Painlev\'e transcendents \cite{Cha11,Cos93,Cos00a,Cos00b,Cos06,Bu39,Bu64a,Bu64b,Bu71}. The Painlev\'e transcendents have long history and they have been obtained by Painleve, Gambier and Fush. These works have been can be found in particular where 50 type of equations whose only movable singularities are poles are presented. The most interesting of the fifty types are those which are irreducible and serve to define new transcendents Painlev\'e transcendents  The other 44 can be integrated in terms of classical functions and transcendents or transformed into the remaining six equations. Several of their properties have been studied over the years in particular of particular solution \cite{Ab78,Co93,Co99,Co08}. They take the following form
  
\begin{subequations}\label{Painleve}
\begin{align}
P_{1}''(z)&=6P_{1}^{2}(z)+z \\ 
P_{2}''(z)&=2P_{2}(z)^{3}+zP_{2}(z)+\alpha \\ 
P_{3}(z)''&=\frac{P_{3}'(z)^{2}}{P_{3}(z)}-\frac{P_{3}'(z)}{z}+\frac{\alpha P_{3}^{2}(z)+\beta}{z}+\gamma P_{3}^{3}(z)+\frac{\delta}{P_{3}(z)}\\ \nonumber
P_{4}^{''}(z)&= \frac{P_{4}^{'2}(z)}{2P_{4}(z)} + \frac{3}{2}P_{4}^{3}(z) + 4zP_{4}^{2}(z) + 2(z^{2} -
\alpha)P_{4}(z) +  \frac{\beta}{P_{4}(z)}\\
P_{5}''(z)&=(\frac{1}{2P_{5}(z)}+\frac{1}{P_{5}(z)-1})P_{5}'(z)^{2}-\frac{1}{z}P_{5}'(z)+\frac{(P_{5}(z)-1)^{2}}{z^{2}}(\frac{aP_{5}^{2}(z)+b}{P_{5}(z)})\\
+&\frac{cP_{5}(z)}{z}+\frac{dP_{5}(z)(P_{5}(z)+1)}{P_{5}(z)-1}\\ \nonumber
P_{6}''(z)&=\frac{1}{2}(\frac{1}{P_{6}(z)}+\frac{1}{P_{6}(z)-1}+\frac{1}{P_{6}(z)-z})P_{6}'(z)^{2}-(\frac{1}{z}+\frac{1}{z-1}+\frac{1}{P_{6}(z)-z})P_{6}'(z)\\ 
+&\frac{P_{6}(z)(P_{6}(z)-1)(P_{6}(z)-z)}{z^{2}(z-1)^{2}}(\gamma_{1}+\frac{\gamma_{2}z}{P_{6}(z)^{2}}+\frac{\gamma_{3}(z-1)}{(P_{6}(z)-1)^{2}}+\frac{\gamma_{4}z(z-1)}{(P_{6}(z)-z)^{2}})\\ \nonumber
\end{align}
\end{subequations}	
	
Many of their properties have been studied in particular particular solutions of $P_{2}$ to $P_{6}$. They find many applications in various domain of mathematical physics Statistical mechanics, quantum field theory, relativity,  Symmetry reduction of various equations (Kdv, Boussineq, Sine-Gordon, Kadomstev-Petviashvile, nonlinear Schrodinger) \cite{Co99}. The connection with non relativistic quantum mechanics and superintegrable systems is much more recent

\subsection{Superintegrable systems with N=3}

Let us consider the case of superintegrable systems with second order integral A and third order integrals B

\begin{equation*}
B=\sum_{i+j+k=3}A_{ijk}\{L_{3}^{i},p_{1}^{j}p_{2}^{k}\}+\{g_{1}(x,y),p_{1}\}+\{g_{2}(x,y),p_{2}\}
\end{equation*}

The constants $A_{ijk}$ and functions V, $g_{1}$ and $g_{2}$ are subject to  determining equation from the commutator $[H,B]=0$

\begin{subequations}\label{det1n3}
\begin{align}
g_{1,x}&=3f_1V_x+f_2V_y\label{det1a}\\
g_{2,y}&=f_3V_x+3f_4V_y\label{det1b}\\
g_{1,y}+g_{2,x}&=2(f_2V_x+f_3V_y) \label{det1c} \\
g_1 V_x+g_2V_y&=\frac{\hbar^{2}}{4}(f_1V_{xxx}+f_2V_{xxy}+f_3V_{xyy}+f_4V_{yyy} )\label{det1d} \\
+&8A_{300}(x_1V_y-x_2V_x)+2(A_{210}V_x+A_{201}V_y)
\end{align}
\end{subequations}

 The functions $f_{i}$ are polynomial involving the constants $A_{ijk}$.
The problem consists in 10 constants and 3 functions to be determined from the overdetermined systems of 4 equations. 20 quantum potential were obtained and among them 5 exotic potentials written in terms of Painlev\'e transcendents. One models in term of the fourth Painlev\'e transcendents can be written as
\newline
\newline
$Q18$
\begin{align*}
V(x,y)=&\frac{\omega^{2}}{2}(x^{2}+y^{2})+\frac{\hbar^{2}}{2}P_{4}^{2}(\sqrt{\frac{\omega}{\hbar}}x)
+2\omega\sqrt{\omega \hbar}P_{4}(\sqrt{\frac{\omega}{\hbar}}x) +\frac{\epsilon\hbar\omega}{2}P_{4}^{'}(\sqrt{\frac{\omega}{\hbar}}x)+\frac{\hbar\omega}{3}(\epsilon-\alpha).\nonumber\\
\end{align*}

\subsection{Cartesian and N=4}

In the case the integrals B is of the fourth order, this integrals can be written in the following form

\begin{equation*}
B=\sum_{j+k+l=4} \frac{A_{jkl}}{2} \{L_3^j,p_1^k p_2^l\}+\frac{1}{2}(\{g_1(x,y),p_1^2\}+\{g_2(x,y),p_1p_2\}+\{g_3(x,y),p_2^2\})+l(x,y)
\end{equation*}

where the quantities $f_i,\; i=1,2,..,5$ are polynomials in the variables x and y. The commutator $[H,B]=0$ lead to a set of 6 linear PDEs for the functions $g_1,g_2,g_3,$ and $l$. We $V$ is not known, this is a system of 6 nonlinear PDEs for $g_i,l$ and $V$. They take the form

\begin{subequations}\label{det1n4}
\begin{align}
g_{1,x}=4f_1V_x+f_2V_y\label{det1a}\\
g_{2,x}+g_{1,y}=3f_2V_x+2f_3V_y\label{det1b}\\
g_{3,x}+g_{2,y}=2f_3V_x+3f_4V_y\label{det1c}\\
g_{3,y}=f_4V_x+4f_5V_y,\label{det1d}
\end{align}
\end{subequations}

and 
\begin{subequations}\label{l1}
\begin{align}
\ell_{x}=&2g_1V_x+g_{2}V_y+\frac{\hbar^2}{4}\bigg((f_2+f_4)V_{xxy}-4(f_1-f_5)V_{xyy}-(f_2+f_4)V_{yyy}\nonumber\\
&+(3f_{2,y}-f_{5,x})V_{xx}-(13f_{1,y}+f_{4,x})V_{xy}-4(f_{2,y}-f_{5,x})V_{yy}\nonumber\\
& -2(6A_{400}x^2+62A_{400}y^2+3A_{301}x-29A_{310}y+9A_{220}+3A_{202})V_x\nonumber\\
& +2(56A_{400}xy-13A_{310}x+13A_{301}y-3A_{211})V_y\bigg),\label{lx}\\
\ell_{y}=&g_{2}V_x+2g_{3}V_y+\frac{\hbar^2}{4}\bigg(-(f_2+f_4)V_{xxx}+4(f_1-f_5)V_{xxy}+(f_2+f_4)V_{xyy}\nonumber\\
&+4(f_{1,y}-f_{4,x})V_{xx}-(f_{2,y}+13f_{5,x})V_{xy}-(f_{1,y}-3f_{4,x})V_{yy}\nonumber\\
&+2(56A_{400}xy-13A_{310}x+13A_{301}y-3A_{211})V_x\nonumber\\
&-2(62A_{400}x^2+6A_{400}y^2+29A_{301}x-3A_{310}y+9A_{202}+3A_{220})V_y\bigg).\label{ly}
\end{align}
\end{subequations}
The quantities $f_i,\; i=1,2,..,5$ are polynomials. The cases can be simplified via transformation such translation and removing
trivial integrals. The leading terms split in 3 types and further analysis will provide 12 exotic cases. For these 12 cases
nonlinear fourth order ordinary differential equation are obtained. They can be further integrated and transformed into an equation of type
Chazy-I. These equations consist in third order differential equations in the polynomial class of the form

\begin{equation}
  W'''=aWW''+bW'^2+cW^2W'+dW^4+A(y)W''
\end{equation}
\[+B(y)WW'+C(y)W'+D(y)W^3+E(y)W^2+F(y)W+G(y),\]

 where $a,b,c,$ and $d$ are certain rational or algebraic numbers, and the remaining coefficients are locally analytic functions of $y$.
Among work on higher order analog \cite{Cha11,Cos93,Cos00a,Cos00b,Cos06,Bu39,Bu64a,Bu64b,Bu71}, the Chazy class \cite{Cha11,Bu64a,Bu64b} has revealed useful in classifying the exotic potential. We relied on the canonical form of Chazy-I equation and its first integral.
In addition, the relation with Painlev\'e transcendent has been facilitated by relying on the following class in which belong the integrals of
Chazy-I

\begin{equation}
A(W',W,y)W''^2+B(W',W,y)W''+C(W',W,y)=0
\end{equation}

where $A$, $B$ and $C$ are polynomials in $W,$ and $W'$  with coefficients analytic in $y$. Cosgrove and Scoufis \cite{Cos93} gave a complete classification of Painlev\'e type equations of second order and second degree $W''^2=F(W',W,y)$ where $F$ is rational in $W',$ and $W$ and analytic in $y$ which divide into six classes of them (denoted by SD-I, SD-II,...,SD-VI). The reduction to Painlev\'e transcendent and transformation were presented. These results allowed to connect for the 12 classes the exotic potential to Painlev\'e trnascendents. One of them is $Q_{3}^{5}$ and given by
\newline
\newline
$Q_3^5:$
\newline
\begin{align*}
V(x,y)=&c_1x+\frac{\hbar ^2}{2}(\sqrt{\alpha } P_3'(y)+\frac{3}{4} \alpha  (P_3(y))^2+\frac{\delta }{4P_3^2(y)}+\frac{\beta P_3(y)}{2y}+\frac{\gamma}{2y P_3(y)}-\frac{ P_3'(y)}{2y P_3(y)}+\frac{P_3'^2(y)}{4 P_3^2(y)}).\nonumber\\
\end{align*}

In addition, a complete classification of doubly exotic potential has been performed for $N=5$ \cite{Ism18}. Another part of that program on the classification of higher order superintegrable systems is the construction of the algebra generated by the conserved quantities.
Among these potentials cases with fourth Painlev\'e transcendents as revealed to lead to interesting algebraic structure formed by the integrals of motion. For more detail in the case N=3 we refer the reader to \cite{mar09a,mar09b}. The explicit form of the integrals have been obtained and involving calculation lead to cubic algebra. The key step consist in using the general formula for the Casimir operator which is a quartic polynomial of the generators and by relying on the explicit differential operator realization to rewritte the Casimir invariant as a polynomial of the Hamiltonian only.  Algebraic expression of the structure function were calculated in regard of the structure constants and finite dimensional unitary representations were obtained.

\section{Constructive approaches}

The works described in previous sections rely on a direct approach to classify the superintegrable systems as well as a direct approach to construct the symmetry algebra. There are advantages to use the direct approach as one exhaust all possibilities and allow to obtain the lowest possible order integrals of motion. On the other hand such approach become harder as the order increase. In recent years different constructive approaches have been proposed where integrals of motion are generate via building block and the corresponding factorized form  facilitate the construction of the symmetry algebra. Among the recent approach are the recurrence and ladder operators method and their classical analog \cite{Cal14a,Kal11,Mar09c,Mar10b,Mar11}. More recently constructive method based on four type of one dimensional operator algebra has been discussed \cite{Saj18}.

\begin{equation}\label{H1}
H_{1}=\frac{p_{x}^{2}}{2}+V(x),\quad K_1=\sum_{j=0}^Mf_j(x)p_x^j
\end{equation}
where $f_M(x) \neq 0,$ and $f_j(x)$ are locally smooth functions.
\newline
Consider the $4$ following forms for $\alpha=\beta=\gamma=0;\; \alpha=\beta=0,\gamma \neq 0;\; \alpha=\gamma=0,\beta \neq 0;$ and $\alpha \neq 0,$ respectively 

\begin{subequations}\label{types}
\begin{equation}
[H_1,K_1]=0,
\end{equation}
\begin{equation}
[H_1,K_1]=\alpha_1,
\end{equation}
\begin{equation}
[H_1,K_1]=\alpha_1 H_1,
\end{equation}
\begin{equation}
[H_1,K_1]=- \alpha_1 K_1,\; \alpha_1 \in \mathbb{R} \backslash 0
\end{equation}
\end{subequations}
where $\alpha_1 \neq 0$ is a constant. We shall refer to these relations as Abelian type (a), Heisenberg type (b), conformal type (c), and ladder type (d), respectively. We shall call the systems $\{H_1,K_1\}$ in one dimension "algebraic Hamiltonian systems". Some of these cases have been studied for particular cases (a) \cite{Hie89,Hie98}, (b) \cite{Fush97,Nik04,Gun14}, (c) \cite{Doe99} and (d) \cite{Ves93,Car04,Mar11,Mar10b}.

\subsection{ Case(d)  }

We consider the one dimensional Hamiltonian and the $M$th order operator $K_1$ and their commutator $[H_1,K_1]$.
Once $[H_1,K_1]$ is chosen to be equal to $0, \alpha_1, \alpha_1H_1$ or $-\alpha_1K_1$, this will provide us with determining equations for the potential $V(x)$ and the coefficients $f_j(x), \; 0\leq j \leq M$ in the operator $K_1.$ We obtain the following operator of order $M+1$

\begin{align}\label{comutDg}
[H_1,K_1]=\sum_{l=0}^{M+1}Z_lD^l
\end{align}
with
\begin{subequations}\label{comutz}
\begin{equation}\label{comutz1}
Z_{M+1}=(-i\hbar )^{M+2} f_M',
\end{equation}
\begin{equation}\label{comutz2}
Z_{M}=-\frac{\hbar ^2}{2}(-i \hbar)^{M-1}(2f_{M-1}'-i\hbar f_{M}''),
\end{equation}
\begin{equation}\label{comutz3}
Z_{l}=-\frac{\hbar ^2}{2}(-i \hbar)^{l-1}(2f_{l-1}'-i\hbar f_{l}'')-\sum_{j=l+1}^{M}(-i\hbar)^{j}f_j C_l^j V^{(j-l)},\; 1 \leq l \leq M-1,
\end{equation}
\begin{equation}\label{comutz4}
Z_0=-\frac{\hbar ^2}{2}f_0''-\sum_{j=1}^{M} (-i\hbar)^{j} f_j V^{(j)}
\end{equation}
\end{subequations}
$C_j^k$ are the Newton binomial coefficients. The determining equations are

\begin{align}
&Z_{M+1}=0,\quad Z_l=\alpha_1 f_l,\quad 0 \leq l \leq M.
\end{align}

The solution to these determining equation gives the following solution for the potentials in terms of Painlev\'e transcendent.

\begin{align}
&V_{d_{1}}=\frac{\alpha_{1}^{2}}{2\hbar ^2} x^{2}, \\\nonumber
&V_{d_2}=\frac{{\alpha_{1}}^2}{8\hbar ^2}x^2+\frac{\beta}{x^2},\\\nonumber
&V_{d_3}=\epsilon \alpha_1 P_4' +\frac{2\alpha_1 ^2}{\hbar ^2}( P_4^{2}+ x P_4)+\frac{\alpha_1 ^2 }{2\hbar ^2}x^{2}+(\epsilon - 1)\frac{\alpha_1}{3}-\frac{\hbar ^2}{6}k_{1},\\\nonumber
&V_{d_4}=\frac{\alpha_1^2}{8\hbar ^2}x^2+\hbar ^2 \big( \frac{\gamma}{P_5-1}+\frac{1}{x^2}(P_5-1)(\sqrt{2\lambda }+\lambda(2P_5-1)+\frac{\beta}{P_5}) \\\nonumber
&+x^2(\frac{P_5'^2}{2 P_5}-\frac{\alpha_1^2}{8\hbar ^4} P_5 )\frac{(2P_5-1)}{(P_5-1)^2}-\frac{P_5'}{P_5-1}-2\sqrt{2\lambda}P_5'\big)+\frac{3\hbar^2}{8x^2}. \\\nonumber
\end{align}

For $M=5$ setting $u(x)=\int{V_{d_{5}}}dx,$ we get the function $f_{i}$ in term of the function u and its derivatives, u satisfies a sixth order differential equation satisfy the Painlev\'e property. These models are interesting from a one dimensional perspective and as illustrated by the Table they allow to generate superintegrable systems. In fact, the corresponding superintegrable systems belonging to $(d,d)$ also lead to polynomial algebras as these ladder operator generate in fact polynomial Heisenberg algebras which can be further use to build the polynomial algebras of the integrals of motion.

\subsection{Case (d,d)}

The operator algebras Case (a), (b), (c) and (d) generate integrals of motion for 

 A 2D system with separation of variables in Cartesian:

\begin{equation*}
  H=H_{1}+H_{2}=-\frac{d^{2}}{dx^{2}}-\frac{d^{2}}{dy^{2}}+V_1(x)+V_2(y)
\end{equation*}

Recently the role played by these operators in context of superintegrability for the classical and quatum models has been
discussed. Let us present part of the construction related to the quantum case.

\begin{table}\label{Tint}
\begin{center}
\begin{tabular}{|l|l|l|l|l|}
  \hline
 Case  &  Type &  Integral type & K  &  Order of $K$  \\[0.05cm]
  \hline	
   &  (b,b)  & polynomial & $\alpha_{2} K_1-\alpha_{1}K_2$ & $max(k_{1},k_{2})$  \\[0.05cm]
  \hline
   &  (c,b) & polynomial &  $\alpha_{2}K_1-\alpha_{1}H_{1}K_2$ & $max(k_{1},k_{2}+2)$      \\[0.05cm]
  \hline
   &  (d,d) & polynomial &  $(K_{1}^{\dagger})^{m} (K_2^{-})^{n} - (K_1^{-})^{m} (K_2^{\dagger})^{n} $ & $(m k_{1}+n k_{2}-1)$     \\[0.05cm]
  \hline	
   &  (c,c) & polynomial &  $\alpha_{2} H_{2}K_1-\alpha_{1} H_{1}K_2$ & $max(k_{1}+2,k_{2}+2)$    \\[0.05cm]
  \hline							
						
\end{tabular}
\caption{\footnotesize{Integrals of motion in $E_2$}}
\end{center}
\end{table}

In the case (d,d) there is existence of polynomial Heisenberg algebras in both axis

\begin{equation*}
  [H_{i},K_{i}^{\dagger}]=\alpha_{i}K{i}^{\dagger},\quad [H_{i},K_{i}^{-}]=-\alpha_{i}K_{i}^{-}
\end{equation*}
\begin{equation*}
 K_{i}^{-}K_{i}^{\dagger}=Q(H_{i}+\alpha_{i}),\quad L_{i}^{\dagger}L_{i}=Q(H_{i})
\end{equation*}

$\alpha_{1}$ and $\alpha_{2}$ are constants while $Q(x)$ and $S(y)$ are polynomials. This is similar to bosonic realization of Lie algebras and in particular boson realizations of $su(2)$ and $su(3)$. Explicit expression for the polynomial algebras $A=H_{1}-H_{2}$ and $B_{1}=(K_{1}^{\dagger})^{m} (K_2^{-})^{n}$, $B_{2}=(K_1^{-})^{m} (K_2^{\dagger})^{n} $ have been obtained and been used to provide spectrum of the quantum models.

\section{Fourth Painlev\'e transcendent model $Q_{18}$: ladder and SUSYQM}

The part of the potential $Q_{18}$ in the x variable and involving the fourth Painlev\'e transcendent has been obtained and rediscovered by different approaches \cite{And00,Car04,Saj18}. In particular, one can generate the model using two types of supercharges in supersymmetric quantum mechanics, of first and second order \cite{And00,Car04}. It has been demonstrated that they imply that the ladder operators can be factorized and this form allow to obtain wavefunctions by determining the zero modes of the annihilation and creation operators. At most three of the six possible states annihilated by $a^-$ ($\psi_{i}$) and $a^+$ ($\phi_{i}$) in total can be square integrable \cite{And00,Car04,mar09b}

\[\psi_{i}=f_{1}(P_{4},P_{4}')e^{\int g_{1}(P_{4},P_{4}')},\quad i=1,2,3\]
\[\phi_{i}=f_{2}(P_{4},P_{4}')e^{\int g_{2}(P_{4},P_{4}')}, \quad i=1,2,3\]

 For some ranges of the parameters $\alpha$ and $\beta$, $H_1$ may admit one, two, or three infinite sequences of equidistant levels, or one infinite sequence of equidistant levels with either one additional singlet or one additional doublet. It has been discussed \cite{mar09b} that these results can be used to compare the spectrum obtained from the cubic algebras for the two dimensional systems which is superintegrable.

\subsection{Example, part 1}

 We considered a case ($\epsilon = 1$) with the two parameter of the fourth Painlev\'e transcendent taking the value $ \alpha = 5$, $\beta = -8$. The fourth Painev\'e admit a rational solution of the form $f(z)=\frac{4z(2z^{2} - 1)(2z^{2}+ 3)}{(2z^{2}+ 1)(4z^{4}+3)}$.

\begin{equation*}
V(x,y)=\frac{\omega^{2}}{2}(x^{2}+y^{2})-\frac{8\hbar^{3}\omega}{(2\omega x^{2}+\hbar)^{2}}+\frac{4\hbar^{2}\omega}{(2\omega x^{2}+\hbar)}+\frac{2\hbar\omega}{3}
\end{equation*}

From the cubic algebra we get unitary representations. Due to constraints for unitary irreducible representations we have one infinite sequence. The spectrum decomposes into an infinite unirreps, a singlet and a doublet.

\begin{subequations}
\begin{align}
\phi(x)&=4\hbar^{4}\omega^{2}x(p+1-x)(x+3)(x+2),\quad E=\hbar \omega(p+\frac{8}{3}),p=0,1,... \\
\phi(x)&=4\hbar^{4}\omega^{2}x(p+1-x)(x-3)(x-1),\quad E=\hbar\omega(p-\frac{1}{3}),p=0 \\
\phi(x)&=4\hbar^{4}\omega^{2}x(p+1-x)(x+1)(x-2),\quad E=\hbar\omega(p+\frac{2}{3}),p=0,1 \\
\end{align}\label{unirreps}
\end{subequations}

Writting in unify way the energie coming from the unitary representations given by (\ref{unirreps}) $E_{N}=\hbar\omega(N-\frac{1}{3})$, with $N=0,1,2,...$. The total degeneracies is thus

\begin{equation}
  \deg(E_{N}) = \begin{cases}
      1  & \text{if $N=0$}, \\ 
      1  & \text{if $N=1$}, \\ 
      1  & \text{if $N=2$}. \\ 
			N-2  & \text{if $N=3,...$}. \\ 
  \end{cases} \label{deg1}
 \end{equation}

Using the ladder operators of third order ( we omit the explicit form here ) and the structure of the three zero modes $a\psi_{0}(x)=0$, $a\chi(x)=0$ and $a^{\dagger}\chi(x)=0$, one can build the whole spectrum in the x-axis by repeated action of the raising operator $\psi_{n}(x)=N_{n}(a^{\dagger})^{n}\psi_{0}$

\begin{equation}
\psi_{0}=e^{\frac{-\omega x^{2}}{2\hbar}}\frac{x(3\hbar +2\omega x^{2})}{(\hbar+2\omega x^{2})}, \chi(x)=\frac{e^{\frac{-\omega x^{2}}{2\hbar}}}{\hbar+2\omega x^{2}}.
\end{equation}

The corresponding spectrum for the two-dimensional models decomposes into the following two infinite sequences of states

\begin{align}
\psi_{n,k}&=\psi_{n}(x)e^{-\frac{\omega y^{2}}{2\hbar}}H_{k}(\sqrt{\frac{\omega}{\hbar}}y),\quad E=\hbar\omega(n+k+\frac{8}{3})\\
\phi_{m}&=\chi(x)e^{-\frac{\omega y^{2}}{2\hbar}}H_{m}(\sqrt{\frac{\omega}{\hbar}}y),\quad E_{m}=\hbar\omega(m-\frac{1}{3})\\
\end{align}

which lead to the following degeneracies 

\begin{equation}
  \deg(E_{N}) = \begin{cases}
      1  & \text{if $N=0$}, \\ 
      1  & \text{for $N=1$}, \\ 
      1  & \text{if $N=2$}. \\ 
			N-1  & \text{if $N=3,...$}. \\ 
  \end{cases} \label{deg2}
 \end{equation}

The total number of degeneracies coming from counting all the states via ladder operators and acting on physical states and given by eq.(\ref{deg2}) is different than the one generated via the cubic algebra formed by integrals of motion and given by eq.(\ref{deg1}). The algebraic approaches allow to get the level, but not the correct degeneracies. This is due to these singlet states, and more precisely to the fact the third order integral obtained is connected to certain type of ladder operators, which is hidden when one classify higher order superintegrable systems. This problem can be observed for other rational solution of the fourth Painlev\'e transcendent and point out how one need to look for additional inner structures and operators for exotic potentials for certain value of their parameters. Here ths particular solutions is connected to Hermite exceptional orthogonal polynomials.

\section{Ladder type (d) , factorization and EOP}

In this section we will provide further details on how the Case (d) of ladder operators is in fact much more richer. It was demonstrated that not only these ladder operators can be factorized, but ladder operators are not unique and even for a same order (i.e. as a differential operators ) distinct lowering and raising operators exist with different patterns of finite and infinite dimensional unitary representations. It means the spectrum can be decomposed into direct sums of distinct finite and infinite unitary representations which belong to different polynomial algebras. In recent years, the case of k-step extension of harmonic oscillator has been related to type III Hermite exceptional orthogonal polynomials. Such orthogonal polynomial has been discovered in 2008, related to quantum models in supersymmetric quantum mechanics and since large body of literature has been devoted to their properties. Let us present some results of SUSYQM. The intertwining of Hamiltonian by two $n$th-order differential operators $\cal A$ and ${\cal A}^{\dagger}$

\begin{equation*}
  {\cal A} H^{(1)} = H^{(2)} {\cal A}, \qquad {\cal A} = A^{(n)} \cdots A^{(2)} A^{(1)}
\end{equation*}
\begin{equation*}
 A^{(i)} = \frac{d}{dx} + W^{(i)}(x),\quad  W^{(i)}(x) = - \frac{d}{dx} \log \varphi^{(i)}(x), i=1, 2, \ldots, n,
\end{equation*}

From seed solution $\varphi_i(x)$ of the Schr\"odinger equation associated with $H^{(1)}$

\begin{equation*}
  \varphi^{(1)}(x) = \varphi_1(x), \qquad \varphi^{(i)}(x) = \frac{{\cal W}(\varphi_1, \varphi_2, \ldots, \varphi_i)}
  {{\cal W}(\varphi_1, \varphi_2, \ldots, \varphi_{i-1})}, i=2, 3, \ldots, n.
\end{equation*}

 Here ${\cal W}(\varphi_1, \varphi_2, \ldots, \varphi_i)$ denotes the Wronskian of $\varphi_1(x)$, $\varphi_2(x)$, \ldots, $\varphi_i(x)$ 

\begin{equation}
  V^{(2)}(x) = V^{(1)}(x) - 2 \frac{d^2}{dx^2} \log {\cal W}(\varphi_1, \varphi_2, \ldots, \varphi_n),
\end{equation}
\begin{equation}
  V^{(1)}(x) = x^2, \qquad - \infty < x < \infty,  
\end{equation}
\begin{equation}
  E^{(2)}_{\nu} = 2\nu + 1, \qquad \nu = -m_k-1, \ldots, -m_2-1, -m_1-1, 0, 1, 2, \ldots,
\end{equation}

State-Adding case : $n=k$ seed functions among the polynomial-type eigenfunctions $\phi_m(x)$ of $H^{(1)}$ below the ground-state energy $E^{(1)}_0$ associated with the eigenvalues $E_m = - 2m -1$. They have the form $\phi_m(x) = {\cal H}_m(x) e^{\frac{1}{2} x^2}$ for $m=0, 1, 2, \ldots$. Thus the chain of seed solution is  $(\varphi_1, \varphi_2, \ldots, \varphi_n)$ is $(\phi_{m_1}, \phi_{m_2}, \ldots, \phi_{m_k})$. The partner potential is nonsingular if $m_1 < m_2 < \cdots < m_k$ with $m_i$ even (resp.\ odd) for $i$ odd (resp.\ even). One central results is the state deleting equivalence i.e. there exist another chain ( not necessarily of the same lenght ) of supercharges in the state deleting approach which generate the same final partner Hamiltonian. In the state deleting approach (at least) $n = m_k + 1 - k$ bound-state wavefunctions of $H^{(1)}$ are used as seed functions: $(\varphi_1, \varphi_2, \ldots, \varphi_n)$ to $(\psi_1, \psi_2, \ldots, \check{\psi}_{m_k-m_{k-1}}, \ldots, \check{\psi}_{m_k-m_2}, \ldots, \check{\psi}_{m_k-m_1}, \ldots, \psi_{m_k})$. The equivalence means

\begin{equation}
V^{(2)}(x) + 2m_k + 2 = \bar{V}^{(2)}(x)
\end{equation}

The $(m_k+1)$th-order differential operator, $c = \bar{\cal A} {\cal A}^{\dagger}$, $c^{\dagger} = {\cal A} \bar{\cal A}^{\dagger}$
a PHA of $m_k$th order, $Q(H^{(2)})$ is indeed a $(m_k+1)$th-order polynomial in $H^{(2)}$

\begin{equation*}
\begin{split}
  & [H^{(2)}, c^{\dagger}] = (2m_k+2) c^{\dagger}, \qquad [H^{(2)}, c] = - (2m_k+2) c, \\
  & [c, c^{\dagger}] = Q(H^{(2)}+2m_k+2) - Q(H^{(2)}), 
\end{split}  \label{eq:PHA}
\end{equation*}
\begin{equation*}
  Q_{ko}(H^{(2)}) = \prod_{i=1}^k (H^{(2)} + 2m_i + 1) \prod_{\substack{j=1 \\ j \ne m_k-m_{k-1}, \ldots, m_k-m_1}}
  ^{m_k} (H^{(2)} - 2j -1),  
\end{equation*}

\section{Families of superintegrable models related to k-step extension}

Many superintegrable families from these k-step rational extension have been constructed \cite{mar13a,mar13b,mar14,mar15,mar16} and in fact these results can be straitforwardly extended in n-dimensional systems in the same way the isotropic and anisotropic harmonic oscillator extend in n dimensional. However, a complete algebraic description of their spectrum would be non trivial. Among these families there are the quantum system

\begin{equation*}
  H= - \frac{d^2}{dx^2} - \frac{d^2}{dy^2} + x^2  + y^2 - 2k - 2 \frac{d^2}{dx^2} 
  \log {\cal W}({\cal H}_{m_1}, {\cal H}_{m_2}, \ldots, {\cal H}_{m_k}),
\end{equation*}
\begin{equation*}
  E_{i,N}=2N, \quad  N=\nu_{x}+\nu_{y}+1
\end{equation*}
\begin{equation*}
 \nu_{x}=-m_{k}-1,\ldots,-m_{1}-1,0,1,2,\ldots, \qquad \nu_{y}=0,1,2\ldots. 
\end{equation*}
 
which display the following spectrum $E_{i,N}=2N$,  $N=\nu_{x}+\nu_{y}+1$ with $\nu_{x}=-m_{k}-1,\ldots,-m_{1}-1,0,1,2,\ldots$, $\nu_{y}=0,1,2\ldots$. The degeneracies take the form

\begin{equation*}
  \deg(E_{N}) = \begin{cases}
      k-j+1  & \text{if $N=-m_{j},-m_j+1,\ldots,-m_{j-1}-1$}, \\ 
        & \text{for $j=2,3,\ldots,k$}, \\ 
      k  & \text{if $N=-m_{1},-m_1+1,\ldots,0$}, \\ 
      N+k  & \text{if $N=1,2,3,\ldots$}. \\ 
  \end{cases} 
 \end{equation*}

This example coincide with the particular solution of the fourth Painlev\'e transcendent we presented in previous section.
In the case $m_{1}=2$ the degeneracies well coincide 

\begin{equation*}
  \deg(E_{N}) = \begin{cases}
      1  & \text{if   $N=-2,-1,0$}, \\ 
      N+1  & \text{if $N=1,2,3,\ldots$}. \\ 
  \end{cases} 
 \end{equation*}

In general these k-step extensions exhibit very unusual
patterns of degeneracies and bands of levels. One application of the ladder operators obtained which only admit infinite dimensional unitary representations is that they allow to build integrals and polynomial algebras whose finite-dimensional unirreps can be combined to provide the correct total number of degeneracies and level. Two approaches have been taken to provide algebraic derivation \cite{mar13a,mar13b,mar14,mar15}, one using the deformed oscillator realizations approach and the other by constructing explicitly in combinatorial ways the multiplets corresponding to reresentations. We classified all multiplets for a given energy level. Let us summarize brieftly this approach. The unirreps may be characterized by $(N,s)$ and their basis states by $ |N, \tau,s, \sigma \rangle$ $\sigma=-s,-s+1,\ldots, s$ and $\tau$ distinguishes between repeated representations specified by the same $s$( integer or half-integer).

\begin{equation*}
  b^{\dagger}|N, \tau,s,s \rangle=b|N, \tau,s,-s \rangle =0. 
\end{equation*}

 The $\sigma$ is associated with each state forming this sequence. Using notation $N= \alpha n_{1}n_{2}+\mu$ with appropriate values of $\alpha$ and $\mu$, $ |N, \nu_{x}  \rangle = |\nu_{x}  \rangle_{1} |N-\nu_{x}-1  \rangle_{2}$. The Tables for the general k-step extended systems, the seven families and generalization to singular oscillator can be find in previous references. Let us only present the case $k=1$ in the Table 2.

\begin{table}[h!]

\caption{}

\begin{center}
\begin{tabular}{lllll}
  \hline\hline\\[-0.2cm]
  $\lambda$ & $\mu$ & $s$ & $\cal N$ & $\deg(E_N)$\\[0.2cm]
  \hline\\[-0.2cm]
  $-1$ & $1, \ldots, m_{1}$ & 0 & 1 & 1 \\[0.2cm]
  0 & 0 & $0$ & 1 & 1  \\[0.2cm]
  $0$& 1, $\ldots,m_{1}$ & $(\frac{1}{2})$ & $\mu$ & $N+1$ \\[0.2cm]
     &  & $0^{\mu-1}$ & & \\[0.2cm] 
  $1,2,\ldots$ & 0 & $(\frac{\lambda}{2})$ & $m_1+1$ & $N+1$ \\[0.2cm]
     & & $(\frac{\lambda-1}{2})^{m_1}$ & & \\[0.2cm]
  $1,2,\ldots$& $1,\ldots,m_{1}$ & $(\frac{\lambda+1}{2})$ & $m_{1}+1$ & $N+1$ \\[0.2cm]
     &  & $(\frac{\lambda}{2})^{\mu-1}$ & & \\[0.2cm]
     &  & $(\frac{\lambda-1}{2})^{m_{1}-\mu+1}$ & & \\[0.2cm]    
  \hline \hline
\end{tabular}
\end{center}

\end{table}

When the parameter $m_{1}=2$, we well recover the total number of degeneracies for the particular case related to the fourth Painlev\'e transcendent given by eq(\ref{deg2}) and thus this provide an algebraic derivation of the spectrum via another set of integrals.

\section{Conclusion}

We presented an overview of the classification of superintegrable quantum systems with a second and an additional third or fourth order integrals, and allowing separation of variable in Cartesian coordinates. In addition to the direct approach to the classification of superintegrable systems, we reviewed results obtained via operator algebras of one-dimensional Hamiltonians. It has been demonstrated that all superintegrable quantum systems are connected to the Painlev\'e property and it is conjectured that it is always the case. The role of the Chazy class of equation has also been highlighted. The conserved quantities of these superintegrable models lead to finitely generated polynomial algebras. For particular values of parameters for models involving the fourth Painlev\'e transcendent an algebraic solution based on integrals generated from a direct approach does not provide an accurate counting of the degeneracies and levels. This was shown by eq(\ref{deg1}) and (\ref{deg2}). An algebric solution was determined using new integrals build from ladder constructed from Darboux-Crum and Krein-Adler supercharges and this is given by Table 2 ( for the specific value $m_{1}=2$ ). This show how algebraic techniques can be extended for higher order superintegrable systems in the case the integrals obtained via the direct approach do not provide a complete solution.

\section*{Acknowledgements}
The research of I.\ M.\ was supported by the Australian Research Council Discovery Project DP 160101376 and a Future Fellowship FT 180100099

\:
\newcommand{\newblock}{}
%\section*{References}


\begin{thebibliography}{99}


\bibitem{Ab78}
Ablowitz M J, Ramani A, and Segur H 1978 Non-linear evolution equations and ordinary differential-equations of
  Painlev\'e type {\it Lett. al Nuo. Cim.} {\bf 23} 333\!\!\!
 
\bibitem{Ism18}
Abouamal I and Winternitz P 2018 Fifth-order superintegrable quantum systems separating in Cartesian coordinates: Doubly exotic potentials 
{\it J. Math. Phys.} {\bf 59} 022104

\bibitem{And00}
Andrianov A, Cannata F, Ioffe M, and Nishnianidze D 2000 
 Systems with higher-order shape invariance: spectral and algebraic properties 
{\it Phys. Lett. A} {\bf 266} 341

\bibitem{Bak1} 
Baker G A 1956 Degeneracy of the n-dimensional, isotropic, harmonic oscillator
{\it Phys. Rev.} {\bf 103} 1119

\bibitem{Ban1} 
Bander M and ItZykson C 1968 Group theory and the hydrogen atom (i)
{\it Rev. Mod. Phys.} {\bf 38} 330

\bibitem{Ba}
Bargmann V 1936 Zur Theorie des Wasserstoffatoms 
{\it Z. Phys.} {\bf 99} 576

\bibitem{Bar1} 
Barut A O 1965 Dynamics of a broken $SU_{N}$ symmetry for the oscillator {\it Phys. Rev.} B {\bf 139} 1433

\bibitem{Bon0}
Bonatsos D , Daskaloyannis C and Kokkotas K 1993 Quantum algebraic description of quantum superintegrable systems in 2 dimensions {\it Phys. Rev.} A {\bf 48} R23407

\bibitem{Bon1} Bonatsos D and Daskaloyannis C 1999 Quantum groups and their applications in nuclear
physics {\it Prog. Part. Nucl. Phys.} {\bf 43} 537

\bibitem{Bud1} 
Budini P 1967 {\it Spectral function sum rules and the decay}
Int. Atomic Energy Agency, Trieste 

\bibitem{Bu39}
Bureau F J 1939 Sur la recherche des \'equations diff\'erentielles du second ordre dont l'int\'egrale g\'en\'erale est \`a points critiques fixes {\it Bull. de la Classe des Sciences} XXV 51

\bibitem{Bu64a}
Bureau F J 1964 Differential equations with fixed critical points
{\it Ann. di Mat. Pura Appl.} LXIV 229

\bibitem{Bu64b}
Bureau F J 1964 Differential equations with fixed critical points
{\it Ann. di Mat. Pura Appl.} LXVI 1

\bibitem{Bu71}
Bureau F J 1971 {\'E}quations diff{\'e}rentielles du second ordre en $Y$ et du second degr{\'e} en $Y''$ dont l'int{\'e}grale g{\'e}n{\'e}rale est {\`a} points critiques fixes {\it Ann. di Mat. Pura Appl.} {\bf 91} 163


\bibitem{Cal14a}
Calzada J A, Kuru S, and Negro J 2014 Superintegrable Lissajous systems on the sphere 
{\it Eur. Phys. J. Plus} {\bf 129} 164


\bibitem{Car04}
Carballo J M, Fern\'andez D J, Negro J and Nieto L M 2004 Polynomial Heisenberg algebras 
{\it J. Phys.} A {\it Math. and Gen.} {\bf 37} 10349

\bibitem{Cha11}
Chazy J 1911 Sur les \'equations diff\'erentielles de troisi{\`e}me ordre et d'ordre sup\'erieur dont l'int\'egrale g\'en\'erale a ses points critiques fixes {\it Acta Mathematica} {\bf 33} 317

\bibitem{Zhe18}
Chen Z, Marquette I and Zhang Y-Z, 2019 Extended Laplace-Runge-Lentz vectors, new family of superintegrable systems and quadratic algebras 
{\it Ann. Phys.}  (to appear)

\bibitem{Co93}
Conte R, Fordy A P, and Pickering A 1993 A perturbative Painlev\'e approach to nonlinear differential
  equations {\it Phys. D: Non. Phen.} {\bf 69} 33

\bibitem{Co99}
Conte R 1999 {\it The Painlev\'e Approach to nonlinear Ordinary Differential Equations. The Painlev\'e property, one century later} 
Springer, New York, 77

\bibitem{Co08}
Conte R and Musette M 2008 {\it The Painlev\'e Handbook}
Springer, Berlin, 810p.

\bibitem{Cos93}
Cosgrove C M and Scoufis G 1993 Painlev\'e classification of a class of differential equations of the second order and second degree 
{\it Stud. in Appl. Math.} {\bf 88} 25
	
\bibitem{Cos00a}
Cosgrove C M 2000 Higher-order Painlev\'e equation in the polynomial class I: Bureau Symbol P2 
{\it Stud. in Appl. Math.} {\bf 104} 1

\bibitem{Cos00b}
Cosgrove C M 2000 Chazy classes IX-XI of third-order differential equations
{\it Stud. in Appl. Math.} {\bf 104} 171

\bibitem{Cos06}
Cosgrove C M 2006 Higher-order Painlev\'e equation in the polynomial class II: Bureau Symbol P1
{\it Stud in Appl Math.} {\bf 116} 321

\bibitem{Das1} Daskaloyannis C 2001 Poisson algebras of two-dimensional classical superintegrable systems and quadratic associative algebras of quantum superintegrable systems {\it J.Math.Phys.} {\bf 42} 1100.

\bibitem{Bie1}
De Bie H, Genest V X, Lemay J-M, Vinet L 2017 A superintegrable model with reflections on $S^{n-1}$ and the higher rank Bannai-Ito algebra {\it J.Phys.} A {\it Math. Theor.} {\bf 50} 195202

\bibitem{Doe99}
Doebner H D and Zhdanov R Z 1999 The stationary KdV hierarchy and so(2,1) as a spectrum generating algebra
{\it J. Math. Phys.} {\bf 40} 4995

\bibitem{Adr17}
Escobar-Ruiz A M and Lopez Vieyra J C and Winternitz P 2017
 Fourth order superintegrable systems separating in polar coordinates. I. Exotic potentials {\it J. Phys.} A {\it Math. Theor.} {\bf 50}  495206

\bibitem{Adr18a}
Escobar-Ruiz A M, Lopez Vieyra J C, Winternitz P and Yurdu\c{s}en I 2018
Fourth-order superintegrable systems separating in Polar Coordinates. II. Standard Potentials {\it J. Phys.} A {\it Math. Theor.} {\bf 51} 455202

\bibitem{Adr18b}
Escobar-Ruiz A M, Winternitz P and Yurdu\c{s}en I 2018
General $N$th order superintegrable systems separating in polar coordinates {\it J. Phys.} A {\it Math.Theor.} {\bf 51} 40LT01

\bibitem{Mil17}
Escobar-Ruiz M A and Miller Jr W 2017 Toward a classification of semidegenerate 3D superintegrable systems {\it J. Phys.} A {\bf 50} 095203

\bibitem{Esc17}
Escobar Ruiz M A, Kalnins E G, Miller W, Subag E 2017 Bocher and Abstract Contractions of 2nd Order Quadratic Algebras {\it SIGMA} {\bf 13} 013

\bibitem{Fo}
Fock V 1935 Zur Theorie des Wasserstoffatoms 
{\it Z. Phys.} A {\bf 98} 145

\bibitem{Fra1} 
Fradkin D M 1965 Three-dimensional isotropic harmonic oscillator and $SU_{3}$ {\it Amer. J. Phys.} {\bf 33} 207 

\bibitem{Win65}
Fri{\v{s}} J, Mandrosov V, Smorodinsky Ya A, Uhl{\'{\i}}{\v{r}} M, and Winternitz P 1965 On higher symmetries in quantum mechanics {\it Phys. Lett.} {\bf 16 } 354

\bibitem{Fuchs84}
Fuchs L I 1884 {\it Uber differentialgleichungen, deren integrale feste verzweigungspunkte besitzen.}
Sitz. Akad. Wiss. Berlin, 699

\bibitem{Fush97}
Fushchych W I and Nikitin A G 1997 Higher symmetries and exact solutions of linear and nonlinear Schr\"{o}dinger equation
{\it J. Math. Phys.} {\bf 38} 5944

\bibitem{Gam10}
Gambier B 1910 Sur les \'equations diff\'erentielles du second ordre et du premier degr\'e dont l'int\'egrale g\'en\'erale est \`a points critiques fixes  {\it Acta Mathematica} {\bf 33} 1

\bibitem{Vin3}
Genest V X, Vinet L and Zhedanov A 2014 Superintegrability in two dimensions and the Racah-Wilson algebra {\it Lett. Math. Phys.} {\bf 104} 931

\bibitem{gomez10a} 
G\'omez-Ullate D, Kamran N, and Milson R 2010 An extension of Bochner's problem: Exceptional invariant subspaces
{\it J. Approx. Theory.} {\bf 162} 987 

\bibitem{gomez09} 
G\'omez-Ullate D, Kamran N, and Milson R 2009 An extended class of orthogonal polynomials defined by a Sturm-Liouville problem
{\it J. Math. Anal. Appl.} {\bf 359} 352

\bibitem{gomez10b} 
G\'omez-Ullate D,  Kamran N, and  Milson R2010 Exceptional orthogonal polynomials and the Darboux transformation
{\it J. Phys.} A {\bf 43} 434016 

\bibitem{ull14}
G\'omez-Ullate D, Grandati Y, and Milson R 2014 Rational extensions of the quantum harmonic oscillator and exceptional Hermite polynomials 
{\it J. Phys.} A {\it Math. Theor.} {\bf 47} 015203 

\bibitem{ull13}
G\'omez-Ullate D, Grandati Y, and Milson R 2014 
Extended Krein-Adler theorem for the translationally shape invariant potentials {\it J. Math. Phys.} {\bf 55} 043510

\bibitem{Gra1}
 Granovskii Ya I, Lutzenko I M and Zhedanov A S 1991 Quadratic algebra as a hidden symmetry of the Hartmann potential {\it J.Phys.} A {\it Math. Gen.} {\bf 24} 3887. 

\bibitem{Gra2}
Granovskii Y, Lutzenko I M and Zhedanov A S 1992 Mutual integrability, quadratic algebras and dynamic symmetry
{\it Ann. Phys.} {\bf 217} 1


\bibitem{Gra02}
Gravel S and Winternitz P 2002 Superintegrability with third-order integrals in quantum and classical mechanics {\it J. Math. Phys} {bf 43} 5902
	
\bibitem{Gra04}
Gravel S 2004 Hamiltonians separable in Cartesian coordinates and third-order integrals of motion {\it J. Math. Phys} {\bf 45} 1003 

\bibitem{Gun14}
Gungor F, Kuru S, Negro J, Nieto L M 2017 Heisenberg-type higher order symmetries of superintegrable systems separable in Cartesian coordinates {\it Nonlin.} {\bf 30} 1788

\bibitem{Gro02} 
Gromak V I, Laine I, and Shimomura S 2002 {\it Painleve Differential Equations in the Complex Plane} Studies in Math {\bf 28} de Gruyter, Berlin, New York, 312p.

\bibitem{Hei15}
Heinonen R, Kalnins E G, Miller Jr W, Subag E 2015 Structure relations and Darboux contractions for 2D 2nd order superintegrable systems {\it SIGMA} {\bf 11} 043

\bibitem{Hie89}
Hietarinta J 1989 Solvability in quantum mechanics and classically superfluous invariants
{\it J. Phys.} A {\it Math. Gen.} {\bf 22} L143

\bibitem{Hie98}
Hietarinta J 1998 Pure quantum integrability
{\it Phys. Lett.} A. {\bf 246} 97

\bibitem{Faz15}
Hoque F, Marquette I, Zhang Y-Z 2015 Quadratic algebra structure and spectrum of a new superintegrable system in N-dimension
 {\it J.Phys.} A {\it Math.Theor.} {\bf 48} 185201 

\bibitem{Hwa1}  
Hwa R C and Nuyts J 1966 Group embedding for the harmonic oscillator
{\it Phys. Rev.} {\bf 145} 1188

\bibitem{In56}
Ince E L 1956 {\it Ordinary differential equations} Dover, New York, 574p.

\bibitem{Isa1}
 Isaac P S and  Marquette I 2014 On realizations of polynomial algebras with three generators
via deformed oscillator algebras {\it J. Phys.} A {\it Math. Theor.} {\bf 47} 205203

\bibitem{Hil}
Jauch J M and Hill E L 1940  On the problem of degeneracy in quantum mechanics
{\it Phys. Rev.} {\bf 57} 641

\bibitem{Kal12}
Kalnins E G and Miller W 2012 Structure results for higher order symmetry algebras of $2D$ classical superintegrable systems
{\it J. Nonlin. Syst. Appl.} {\bf 3} 29

\bibitem{Kal11} 
Kalnins E G, Kress J M, and Miller W 2011 A recurrence relation approach to higher order quantum superintegrability
{\it SIGMA} {\bf 7} 031

\bibitem{Kal13}
Kalnins E G, Miller Jr W, Post S 2013 Contractions of 2D 2nd order quantum superintegrable systems and the Askey Scheme for hypergeometric orthogonal polynomials {\it SIGMA} {\bf 9} 057 

\bibitem{Li18}
Liao Y, Marquette I, Zhang Y-Z 2018 Quantum superintegrable system with a novel chain structure of quadratic algebras
{\it J. Phys.} A {\it Math. Theor.} {\bf 51} 255201 

\bibitem{Lou1} 
Louck J D and Galbraith H W 1972 Application of orthogonal and unitary group methods to the n-body problem
{\it Rev. Mod. Phys.} {\bf 44} 540

\bibitem{Lou2} 
Louck J D Group theory of harmonic oscillators in n-dimensional space {\it J. Math. Phys.} {\bf 6} 1786

\bibitem{Sno15}
Marchesiello A, Post S and Snobl L 2015 Third-order superintegrable systems with potentials satisfying only nonlinear equations
{\it J. Math. Phys.} {\bf 56} 102104

\bibitem{Saj17}
Marquette I, Sajedi M and Winternitz P 2017 Fourth order superintegrable systems separating in Cartesian coordinates I. Exotic quantum potentials {\it J. Phys. } A {\it Math. Theor.} {\bf 50} 315201

\bibitem{Saj18}
Marquette I, Sajedi M and Winternitz P 2019 Two-dimensional superintegrable systems from operator algebras in one dimension {\it J. Phys.} A {\it Math. Theor.} (to appear)

\bibitem{Mar18}
Marquette I and Winternitz P {\it Another Painlev\'e conjecture}, Integrability, Supersymmetry and Coherent States A volume in honour of Professor V\'eronique Hussin ( in preparation )

\bibitem{mar15}
Marquette I 2015 {\it New families of superintegrable systems from k-step rational extensions, polynomial algebras and degeneracies}
J. Phys.: Conf. Ser. {\bf 597} 012057

\bibitem{Que15}
Marquette I and Quesne C 2015 Deformed oscillator algebra approach of some quantum superintegrable Lissajous systems on the sphere and of their rational extensions {\it J. Math. Phys.} {\bf 56} 062102 

\bibitem{mar09a}
Marquette I 2009 Superintegrability with third order integrals of motion, cubic algebras, and supersymmetric quantum mechanics. I. Rational function potentials {\it J. Math. Phys.} {\bf 50} 012101

\bibitem{mar09b}
Marquette I 2009 Superintegrability with third order integrals of motion, cubic algebras, and supersymmetric quantum mechanics. II. Painlev\'e transcendent potentials {\it J. Math. Phys} {\bf 50} 095202

\bibitem{mar12}
Marquette I 2013 Quartic Poisson algebras and quartic associative algebras and realizations as deformed oscillator algebras {\it J. Math. Phys.} {\bf 54} 071702 


\bibitem{Mar09c} 
Marquette I 2009 Supersymmetry as a method of obtaining new superintegrable systems with higher order integrals of motion
{\it J. Math. Phys.} {\bf 50} 122102

\bibitem{Mar10b}
Marquette I 2010 Construction of classical superintegrable systems with higher order integrals of motion from ladder operators
{\it J. Math. Phys.} {\bf 51} 072903

\bibitem{Mar11}
Marquette I 2011 {\it An infinite family of superintegrable systems from higher order ladder operators and supersymmetry}
J. Phys.: Conf. Ser. {\bf 284} 012047


\bibitem{mar13a} 
Marquette I and Quesne C 2013 Two-step rational extensions of the harmonic oscillator: exceptional orthogonal polynomials and ladder operators {\it J. Phys.} A {\it Math. Theor.} {\bf 46} 155201 

\bibitem{mar13b}
Marquette I and Quesne C 2013 New ladder operators for a rational extension of the harmonic oscillator and superintegrability of some two-dimensional systems {\it J. Math. Phys.} {\bf 54} 102102

\bibitem{mar14}
Marquette I and Quesne C 2014 Combined state-adding and state-deleting approaches to type III multi-step rationally extended potentials: Applications to ladder operators and superintegrability {\it J. Math. Phys.} {\bf 55} 112103

\bibitem{mar16}
Marquette I and Quesne C 2016 Connection between quantum systems involving the fourth Painlev\'e transcendent and 
k-step rational extensions of the harmonic oscillator related to Hermite EOP {\it J. Math. Phys.} {\bf 57} 052101


\bibitem{Mil13}
Miller W, Post S, and Winternitz P 2013 Classical and quantum superintegrability with applications
{\it J. Phys.} A {\bf 46} 423001

\bibitem{Mos1} 
Moshinsky M 1962 Wigner coeffients for the su(3) group and some applications {\it Rev. Mod. Phys.} {\bf 34} 813


\bibitem{Nik04}
Nikitin A G 2004 {\it Higher-order symmetry operators for Schr\"{o}dinger equation} CRM Proceedings and Lecture Notes {\bf 37} (AMS)

\bibitem{oda11}
Odake S and Sasaki R 2011 Exactly solvable quantum mechanics and infinite families of multi-indexed orthogonal polynomials
{\it Phys. Lett.} B {\bf 702} 164 

\bibitem{oda13}
Odake S and Sasaki R 2013 Krein-Adler transformations for shape-invariant potentials and pseudo virtual states
{\it J. Phys.} A {\it Math. Theor.} {\bf 46} 245201 

\bibitem{Pa02}
Painlev\'e P 1902 Sur les \'equations diff\'erentielles du second ordre et d'ordre sup\'erieur dont l'int\'egrale g\'en\'erale est uniforme {\it Acta Mathematica} {\bf 25} 1

\bibitem{Pop12}
Popper I, Post S, Winternitz P 2012 Third-order superintegrable systems separable in parabolic coordinates {\it J. Math. Phys.} {\bf 53} 062105

\bibitem{Pos15}
Post S and Winternitz P 2015 General Nth order integrals of motion in the Euclidean plane {\it J. Phys.} A {\it Math. Theor.} {\bf 48} 405201	

\bibitem{cq09} 
Quesne C 2009 Solvable rational potentials and exceptional orthogonal polynomials in supersymmetric quantum mechanics
{\it SIGMA} {\bf 5} 084

\bibitem{Ras1} 
Rasmussen W O and Salamo S 1979 An algebraic approach to Coulomb  {\it J. Math. Phys.} {\bf 20} 1064 

\bibitem{Sud1} 
Sudarshan E C G , Mukunda N and O'Raifeartaigh L 1965 Group theory of the Kepler problem
{\it Phys. Lett.} {\bf 19} 322


\bibitem{Tan1} Tanoudis Y and Daskaloyannis C 2011 Algebraic calculation of the energy eigenvalues for the nondegenerate three-dimensional Kepler-Coulomb potential {\it SIGMA} {\bf 7} 054

\bibitem{Tre10}
Tremblay F and Winternitz P 2010 Third-order superintegrable systems separating in polar coordinates {\it J. Phys.} A  {\bf 43} 175206

\bibitem{Ves93}
Veselov A P and Shabat A 1993 Dressing Chains and the Spectral Theory of the Schr\"{o}dinger Operator 
{\it Funkc. Anal. Priloz.} {\bf 27} 1 



\end{thebibliography}
\end{document}